# Density estimation of weak periodic signals in pre-earthquake seismic waves

Nazmi Yılmaz[1,]
[1]*Koç University, College of Sciences, Department of Physics, Istanbul, Turkiye*
*nayilmaz@ku.edu.tr*
*https://orcid.org/0000-0002-0631-257X*

## Abstract

We introduce a method for identifying weak periodic components in pre-earthquake seismic waveforms by examining the scale-index response of a driven Duffing chaotic oscillator. This nonlinear setup helps detect and classify subtle deterministic features buried in low-amplitude, noisy seismic records. We apply this approach to seismic data collected before three moderate-to-strong earthquakes, and compare the results with a quiescent control period. The weak periodic signals detected with the approach exhibit clear, systematic shifts in frequency. Kernel density estimates highlight these changes in the dynamics of the Marmara fault segment south of Istanbul, the likely location of a future large Istanbul earthquake. The results indicate also that chaos-based detection methods can reveal possible precursory patterns that point to the potential of such approaches for real-time seismic monitoring and forecasting. More generally, this study adds to the growing interaction between nonlinear dynamics and geophysics by providing another perspective on the complex behaviour of active fault systems. The density-based analysis of weak periodic signals proposed here may also be useful in other fields that involve large, noisy datasets with chaotic characteristics, such as physiology or financial systems.

## 1. Introduction

Weak periodic or quasi-periodic signals are regular oscillations buried within the stochastic fluctuations of dynamical systems. They are often hidden in the given signal due to their extremely low signal-to-noise ratios (SNR). In many cases, the signal amplitude is so small compared to the noise that standard linear detection methods are not capable to identify them[1,2]. However, these weak components are not random; they carry deterministic information about the system's internal dynamics. Therefore, it is still possible to extract these weak periodic signals from noisiy non-stationary data using nonlinear approaches that exploit the underlying periodic behaviour [3–8].

While methods such as Stochastic Resonance or Chaotic Resonance use noise-assisted amplification to reveal a hidden periodic signal, they often require careful parameter adjustments or even the target waveform reshaping [9–15]. The Duffing oscillator, also sometimes used with other oscillators, including the Van der Pol oscillator, offers clear alternative for detecting weak periodics signals in chaotic time series [16–19]. When the Duffing oscillator is set close to the edge of phase-transition, even a weak periodic input can shift the system from chaotic to a periodic state, effectively acting as a sensitive indicator for the presence of weak periodic signals [20–22]. To detect this shift, we use a wavelet-based scale index that ranges from 0 to 1: periodic motion drives the index toward zero, meantime chaotic dynamics keep it at positive values. Then, plotting the scale index versus frequency graph allows automated extraction of weak periodic signals [23–27].

In seismic wave analysis, detecting underlying weak periodic signals is crucial as they provide insights into fault behavior and dynamics. In this study, we apply an automated, scale-index-based Duffing oscillator method to low-amplitude seismic records from the North Anatolian strike-slip fault zone. The data includes three moderate-to-strong shallow earthquakes that occurred near Istanbul [28], along with a control period during which no significant seismic activity was present. Detecting such weak signatures is particularly relevant given the 30-year probability of an $M > 7.3$ earthquake in the region, which rises to about 47% when the stress transferred by the 1999 Izmit event is taken into account [29]. Another recent work also points to creeping behaviour in the Central Basin segment, adjacent to the locked Kumburgaz Basin section of the Marmara fault south of the city [30].

We compute the scale-index of the numerical solutions of the Duffing chaotic oscillator to extract weak periodic signlas in seismic waves. We then use Kernel density estimation (KDE) to obtain the density of the weak signal frequency spectrum in the seismic waves. This approach proves more effective than traditional phase-space techniques. It also differs from stochastic resonance and chaotic resonance which need noise optimization or the simplification of target waveforms into pure sinusoids. In this work, we also improve this Scale index-Duffing based automated WPS detection method, using a single scan to extract density information across a wide frequency spectrum from multiple low amplitude seismic wave time series. Consequently, our enhanced nonlinear methodology proves computationally efficient and applicable for real-time seismic monitoring. It also allows deeper physical understanding of the precursory periodic features within complex seismic systems, with potential applications in other domains such as physiology and econometry.

## 2. Methodology

### 2.1 Theoretical Basis: The Nonlinear Duffing System

In this study, we use nonlinear Duffing oscillator to extract weak periodic signals embedded in seismic wave time series. The dimensionless Duffing equation is expressed as [3, 6, 31-33];

$$\ddot{x} + \delta\dot{x} - x + x^3 = \gamma\cos(t) + input(t) \quad (1)$$

where $\delta$ is the damping parameter, $-x + x^3$ represents the nonlinear restoring force, and $\gamma\cos(\omega t)$ is the harmonic driving term with amplitude $\gamma$ and angular frequency $\omega$. The external *input* term corresponds to the seismic time series, embedded to the Duffic oscillator.

To enable wide spectrum frequency scanning, we rescale time via $t = \omega_0\tau$, where $\omega_0$ denotes the reference scanning frequency. The equation of motion becomes

$$\dot{x} = \omega_0 y,\ y = \dot{x},\ \dot{y} = \omega_0\big(-0.5y + x - x^3 + \gamma\cos(\omega_0\tau) + \mathrm{input}(\tau)\big) \quad (2)$$

For fixed damping parameter $\delta$, the driving amplitude $\gamma$ determines the behaviour of the oscillator. As $\gamma$ increases, the Duffing oscillator undergoes a transition from low-amplitude periodic state to chaos and eventually to a large scale periodicity at a critical threshold $\gamma_c$. For $\gamma < \gamma_c$ the system exhibits chaotic behaviour, whereas for $\gamma \geq \gamma_c$ the oscillator settles into a stable periodic state**.** This transition at $\gamma_c$ forms the basis of the weak-signal detection mechanism used in this study. We performed all simulations using a fourth-order Runge–Kutta numerical integrator [34].

### 2.2 Detection Mechanism via Chaos-Threshold Tuning

Signal detection exploits the system's sensitivity at the chaos–periodicity boundary. The oscillator amplitude is adjusted close to its critical driving value ($\gamma \approx \gamma_c$), the presence of a weak periodic component in $input$ at frequency $\omega$ perturbes the system and the chaotic attractor collapses into a periodic state, indicating that the weak signal of a certain frequency has been detected, even at extremely low SNR.

### 2.3 Automated Identification: Wavelet-Based Scale Index

Conventional phase-space analysis methods are impractical for continuous seismic monitoring due to their computational cost [6]. For automated signal extraction, we use the scale index ($i_{\text{scale}}$) which is an indicator of aperiodicity derived from Continuous Wavelet Transform (CWT) [23, 24].

The CWT of a signal $f(t)$is;

$$W_f(u,s) = \langle f, \psi_{u,s} \rangle = \int_{-\infty}^{+\infty} f(t)\ \psi_{u,s}^*(t)\, dt \tag{3}$$

The wavelet scalogram is expressed using CVT for a given scale interval;

$$S(s) = \|Wf(u,s)\| = \left(\int_{-\infty}^{+\infty} |Wf(u,s)|^2 du\right)^{1/2} \tag{4}$$

For finite time series, the inner wavelet scalogram is expressed as;

$$S^{inner}(s) = \|Wf(u,s)\|_{J(s)} = \left(\int_{c(s)}^{d(s)} |Wf(u,s)|^2\, du\right)^{1/2} \tag{5}$$

Here $J(s)$ is the sub-interval of maximum length in the given time interval that the one can apply a wavelet function. Hence, the wavelet scale index for a signal $f$ , is computed from the normalised inner scalogram [24];

$$i_{scale} = \frac{\bar{s}^{inner}(s_{min})}{\bar{s}^{inner}(s_{max})} \tag{6}$$

Where $s_{\text{min}}$ and $s_{\text{max}}$ denote the minimum and maximum energy scales in a given scale interval. The computed wavelet scale index provides a clear diagnosticof the systems chatioc state: values of $i_{\text{scale}} \to 0$ indicate periodic or quasi-periodic behaviour and are interpreted as successful detections of weak embedded signals, whereas values of $i_{\text{scale}} \gg 0$ correspond to aperiodic or chaotic dynamics, signifying the absence of coherent periodic components in the seismic record [24, 35-37].

### 2.4 Implementation for Seismic Wave Analysis

The WPS detection method in seismic time series consists of four sequential stages. First, the Duffing oscillator parameters are configured by fixing the damping parameterat $\delta = 0.5$ and tuning the oscillator amplitude to the chaos–periodicity threshold. Next, the seismic *input* term is empirically scaled via a gain coefficient to ensure that its amplitude lies within the operational range of the oscillator. Then, a frequency-sweeping is performed, in which the system is scanned across a predetermined interval $[f_{\text{min}}, f_{\text{max}}]$. The Duffing equation is numerically solved for each frequency, and the scale index $i_{\text{scale}}$ is calculated from the resulting trajectory $x(t)$. Finally, weak signals are extracted by determining the frequencies at which the scale index falls below the threshold $i_{\text{scale}} < 10^{-2}$, indicating that the Duffing oscillator shifts from chaotic behaviour to periodicity in response to the embedded weak signal.

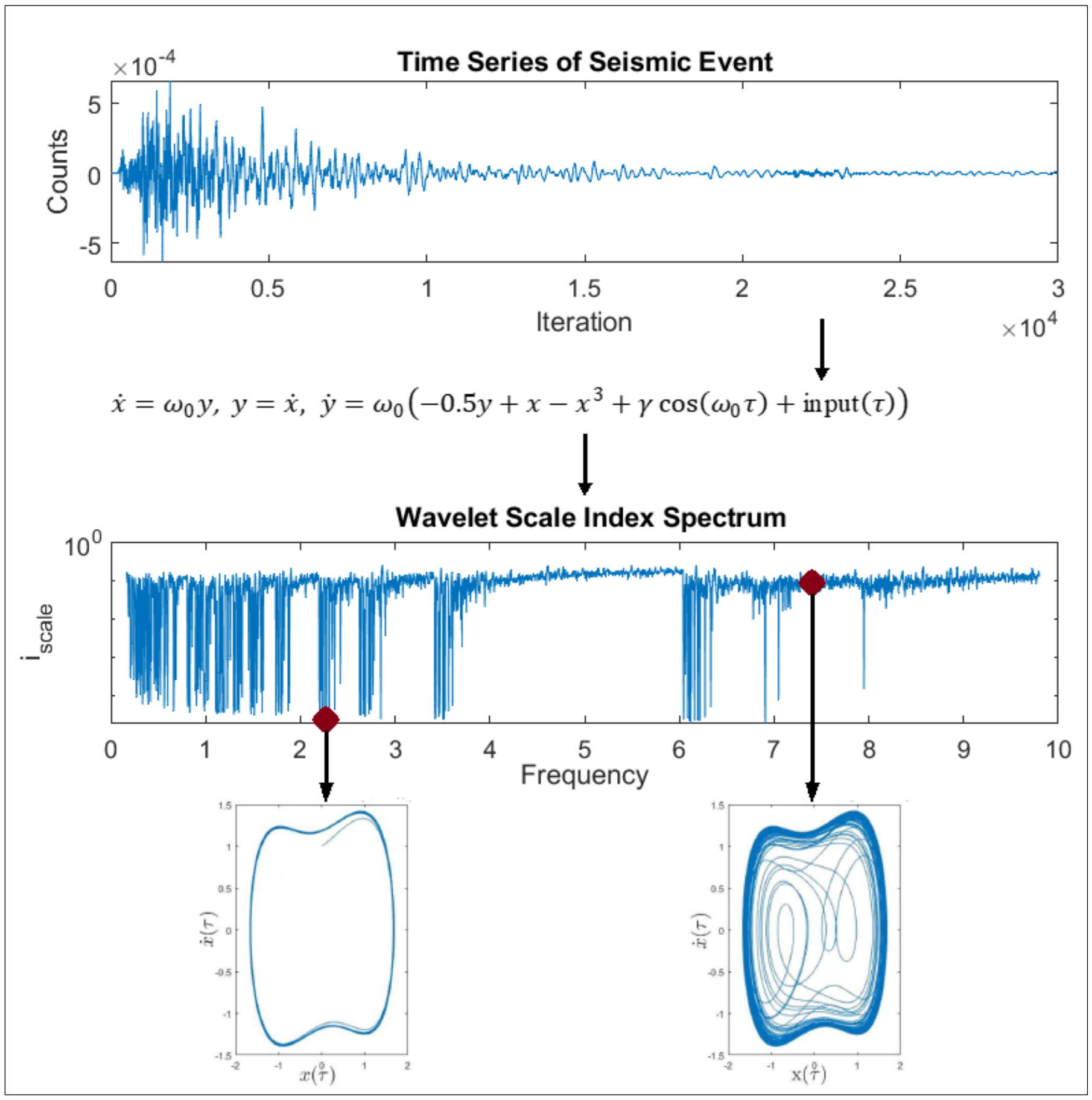

**Figure 1.** The analytical methodology used for automated weak periodic signal detection in seismic recordings: The Duffing oscillator is driven by seismic input, wavelet scale index spectrum is obtained and WPS in a determined frequency range is detected.

## 3. Seismic Data and Study Area

### 3.1 Data Acquisition and Selection

We obtained the seismic data from the Boğaziçi University Kandilli Observatory and Earthquake Research Institute (KOERI) [28]. To investigate the evolution of periodic features, we extracted five-minute velocity–time windows at specific intervals: 5-0 minutes, 1, 2, 3 hour(s) and 5-0 prior to the onset of each seismic event, as well as during the mainshock phase.

For statistical control, five independent five-minute segments were selected from a 5-hour period of seismic inactivity on 27 December 2024. To ensure comparability with the earthquake recordings, these inactive control windows were chosen from similar local times of day, minimizing the effect of temporal variations in microseisms and anthropogenic noise.

All four datasets, the three earthquake events and the inactive control period, were captured at the Silivri broadband seismic station (41.23° N, 28.21° E), aproximately 40 km north of the NAF segments under investigation. Using a single station ensures that differences in dynamical behavior arise from the seismic source processes rather than station-specific or environmental variability.

### 3.2 Seismotectonic Context

The analysis concentrates on three moderate-to-strong earthquake events that accored along the Main Marmara Fault as shown in Figure 2. The first event, Istanbul-1 with $M_w = 5.7$ and depth of 12.6 km occured at 40.88 N, 28.11 E on 26 September 2019. The second event, Istanbul-2 with $M_w 6.2$ and depth of 12.1 km originated at 40.84 N, 28.23 E on 23 April 2025, reflecting a significant strain-release episode within the Kumburgaz Basin/Segment [24]. The third event, Istanbul-3 with $M_w 5.0$ and depth of 12.5 km occured at 40.80 N, 27.95 E on 02 October 2025. For the purpose of statistical comparision and for establishing a baseline response of the nonlinear detection system, we also added a fourth dataset, an inactive control period recorded on 27 December 2024 at Silivri station.

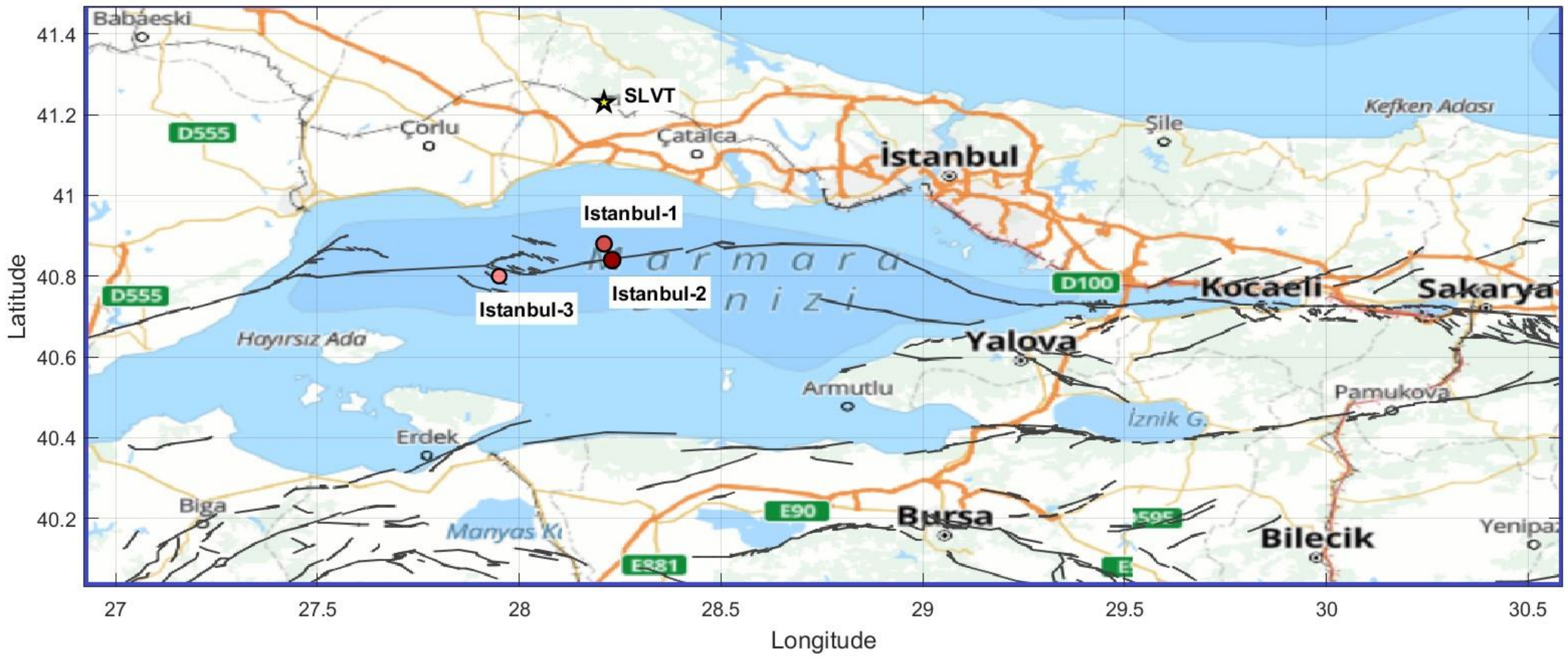

**Figure 2.** The locations of Silivri seismic station (SLVT) and the events of Istanbul-1, 26 September 2019 (magnitude 5.7), Istanbul-2, 23 April 2025 (magnitude 6.2) and Istanbul-3, 02 October 2025 (magnitude 5.0) shown in the Marmara Sea map. The black lines show the NAF segments along the region [28].

## 4. Results

### 4.1 Frequency Scanning and Parameter Calibration

We performed systematic frequency scanning of the Duffing oscillator solutions across the angular frequency ω interval from 0.052 to 3.078 corresponding to a seismic frequency interval of 0.166 Hz to 9.80 Hz. This wide frequency range sweep ensures the capture of both low and high frequency seismic acticity.

The bifurcation threshold $\gamma_c$ of the oscillator is frequency dependent as the driving amplitude $\gamma$ required to trigger a state transition increases with $\omega$. The amplitude $\gamma$ values dynamically scaled for each frequency sub-band to maintain the oscillator at the critical onset of chaos, are summarized in Table 1.

Table 1: The amplitude values of the seismic input for the oscillator in the interval of $0.052 \leq \omega_0 \leq 3.078$.

| $\omega_0$ range | $\gamma$ | $\gamma_c$ |
|---|---|---|
| $0.052 \leq w_0 \leq 0.057$ | 0.82525 | 0.82530 |
| $0.058 \leq w_0 \leq 0.064$ | 0.82530 | 0.82535 |
| $0.065 \leq w_0 \leq 0.075$ | 0.82535 | 0.82540 |
| $0.076 \leq w_0 \leq 0.081$ | 0.82540 | 0.82542 |
| $0.082 \leq w_0 \leq 0.085$ | 0.82542 | 0.82543 |
| $0.086 \leq w_0 \leq 0.089$ | 0.82543 | 0.82544 |
| $0.090 \leq w_0 \leq 0.094$ | 0.82544 | 0.82545 |
| $0.095 \leq w_0 \leq 0.098$ | 0.82545 | 0.82546 |
| $0.099 \leq w_0 \leq 0.103$ | 0.82546 | 0.82547 |
| $0.104 \leq w_0 \leq 0.111$ | 0.82547 | 0.82548 |
| $0.112 \leq w_0 \leq 0.119$ | 0.82548 | 0.82549 |
| $0.120 \leq w_0 \leq 0.128$ | 0.82549 | 0.82550 |
| $0.129 \leq w_0 \leq 0.141$ | 0.82550 | 0.82551 |
| $0.142 \leq w_0 \leq 0.154$ | 0.82551 | 0.82552 |
| $0.155 \leq w_0 \leq 0.176$ | 0.82552 | 0.85553 |
| $0.177 \leq w_0 \leq 0.205$ | 0.82553 | 0.82554 |
| $0.206 \leq w_0 \leq 0.253$ | 0.82554 | 0.82555 |
| $0.254 \leq w_0 \leq 0.283$ | 0.825550 | 0.825554 |
| $0.280 \leq w_0 \leq 0.307$ | 0.825554 | 0.825557 |
| $0.308 \leq w_0 \leq 0.347$ | 0.825557 | 0.825561 |
| $0.348 \leq w_0 \leq 0.406$ | 0.8255561 | 0.8255563 |
| $0.407 \leq w_0 \leq 0.461$ | 0.825563 | 0.825565 |
| $0.462 \leq w_0 \leq 0.545$ | 0.825565 | 0.825567 |
| $0.546 \leq w_0 \leq 0.690$ | 0.825567 | 0.825569 |
| $0.691 \leq w_0 \leq 0.823$ | 0.825569 | 0.82557 |
| $0.824 \leq w_0 \leq 1.073$ | 0.82557 | 0.825571 |
| $1.074 \leq w_0 \leq 1.897$ | 0.825571 | 0.825572 |
| $1.899 \leq w_0 \leq 2.128$ | 0.8255720 | 0.8255721 |
| $2.129 \leq w_0 \leq 2.479$ | 0.8255721 | 0.8255722 |
| $2.480 \leq w_0 \leq 3.078$ | 0.8255722 | 0.8255723 |

### 4.2 Scale Index Computation

For each discrete frequency step within the given range, the numerical solution of the Duffing oscillator was obtained $x(t)$. The scale index $i_{scale}$ was subsequently derived from the resulting time series to quantify the degree of aperiodicity with $i_{scale}$ converging to zero being the primary diagnostic for identifying weak signals. By plotting the scale index as a function of frequency, we generated a high-resolution spectral map that highlights specific detection windows where the seismic input successfully drove the oscillator into a periodic regime.

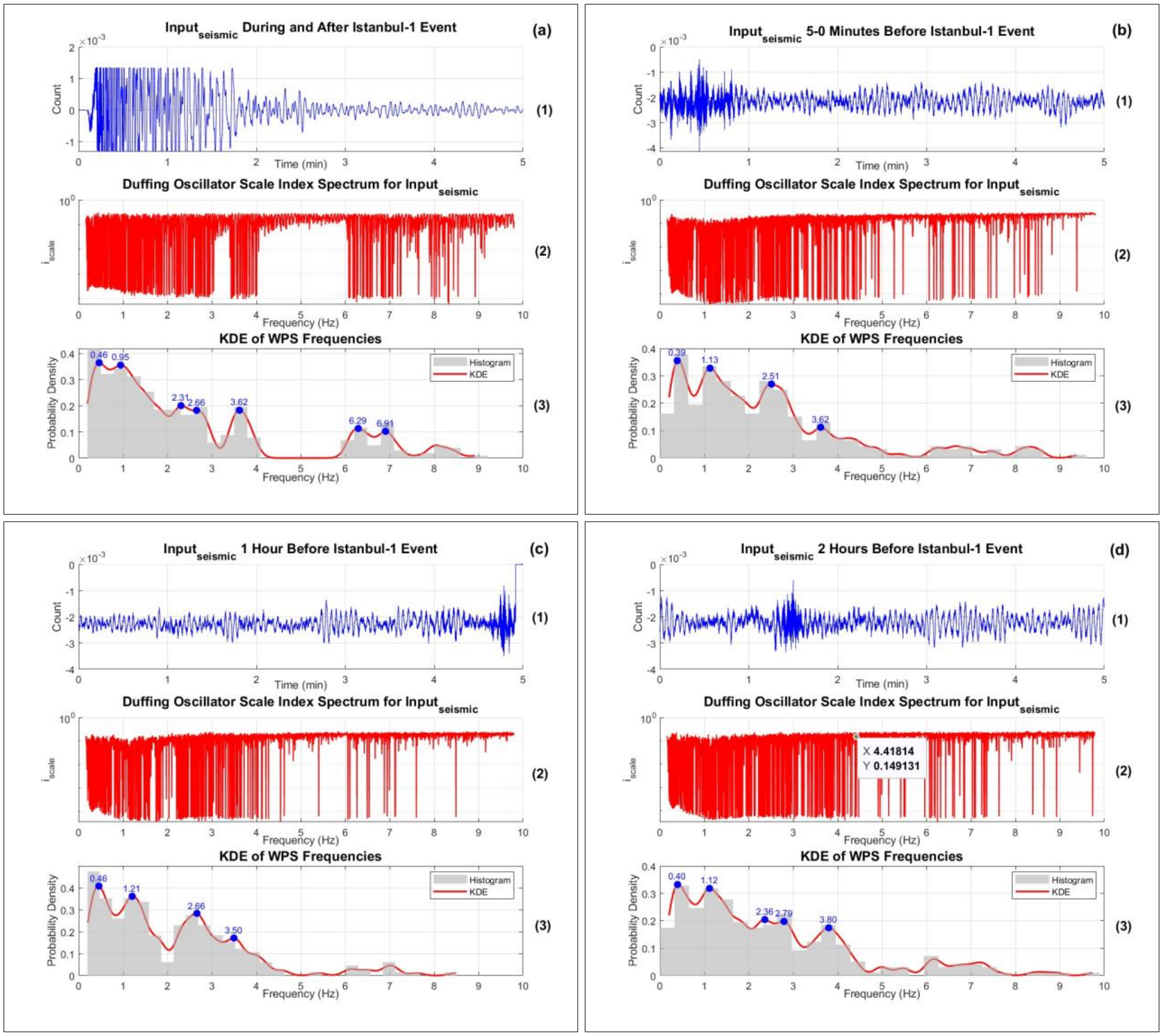

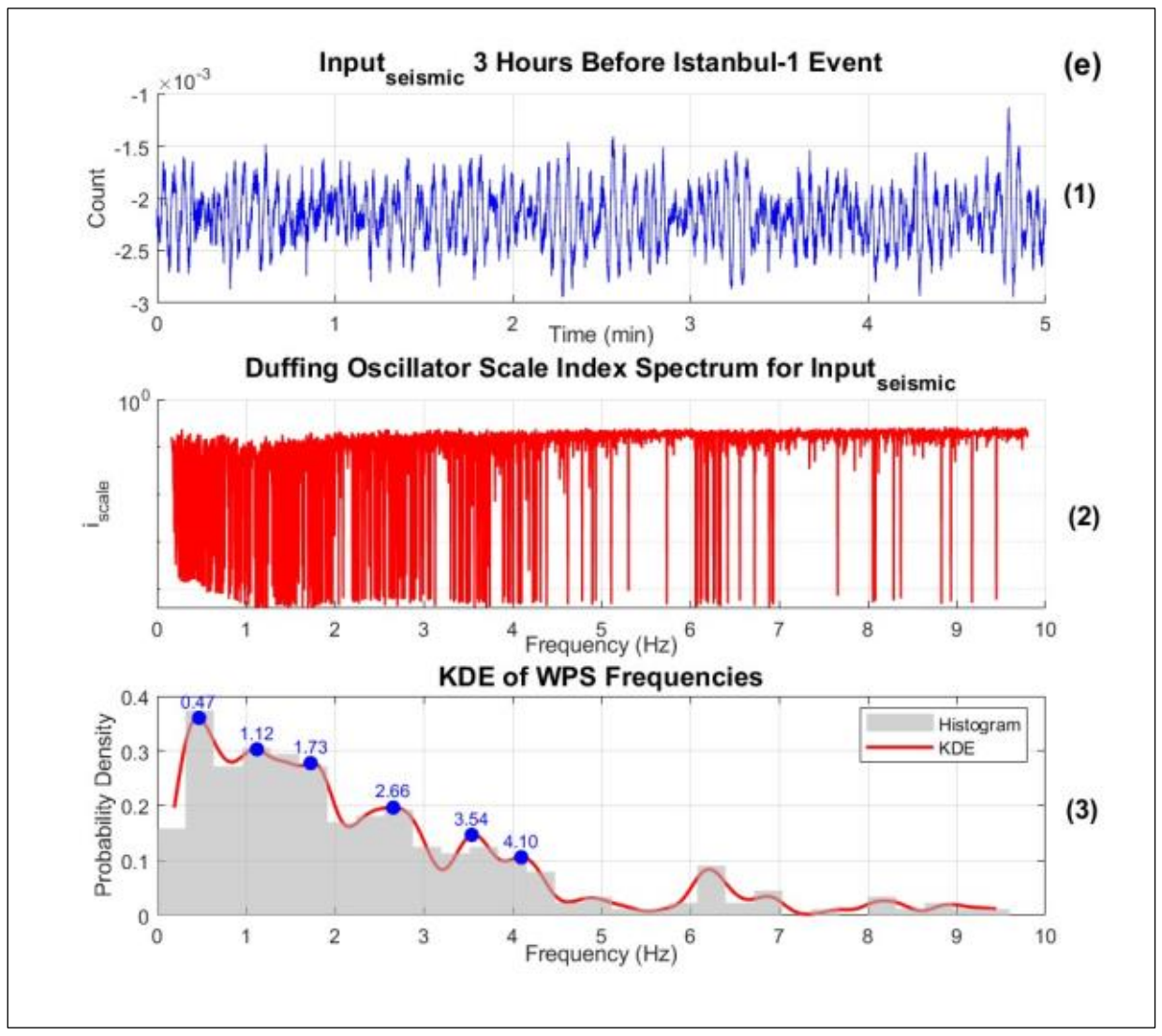

**Figure 3:** Coparison of seismic input, Duffing oscillator wavelet scale index spectrum and WPS frequency distributions across five time windows relative to the Istanbul-1 event. Panels **(a–e)** correspond to: **(a)** during/after the event (5-min), **(b)** 0–5 min before, **(c)** 1 hour before, **(d)** 2 hours before, and **(e)** 3 hours before. For each window, subplots 1–3 show: **(1)** Seismic wave time-series, **(2)** Duffing oscillator wavelet scale index spectrum, and **(3)** KDE of WPS peak frequencies. Together, the 15 subplots **(a1–e3)** illustrate the evolution of seismic signals, nonlinear spectral behaviour, and dominant frequency content leading up to and during the event.

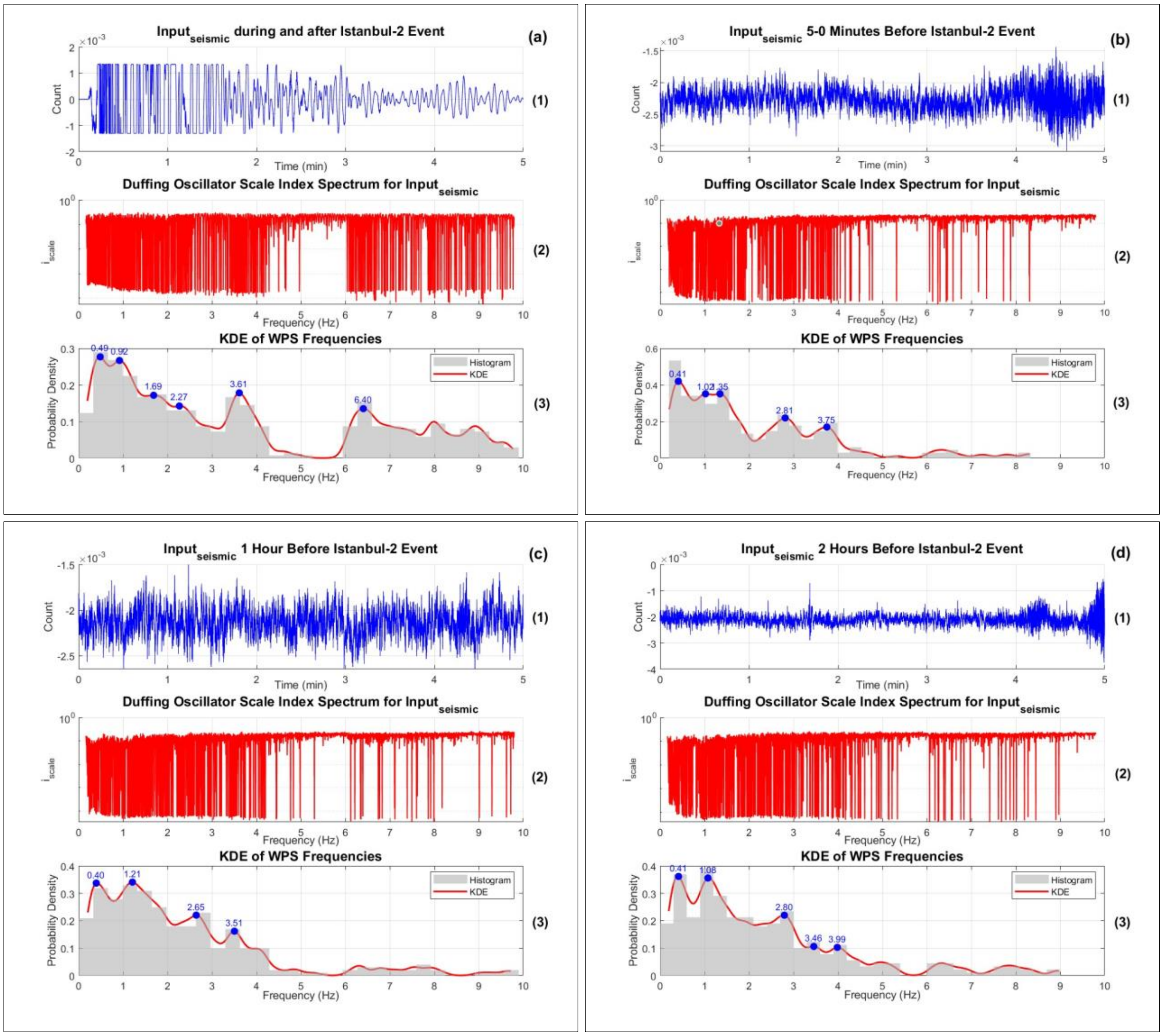

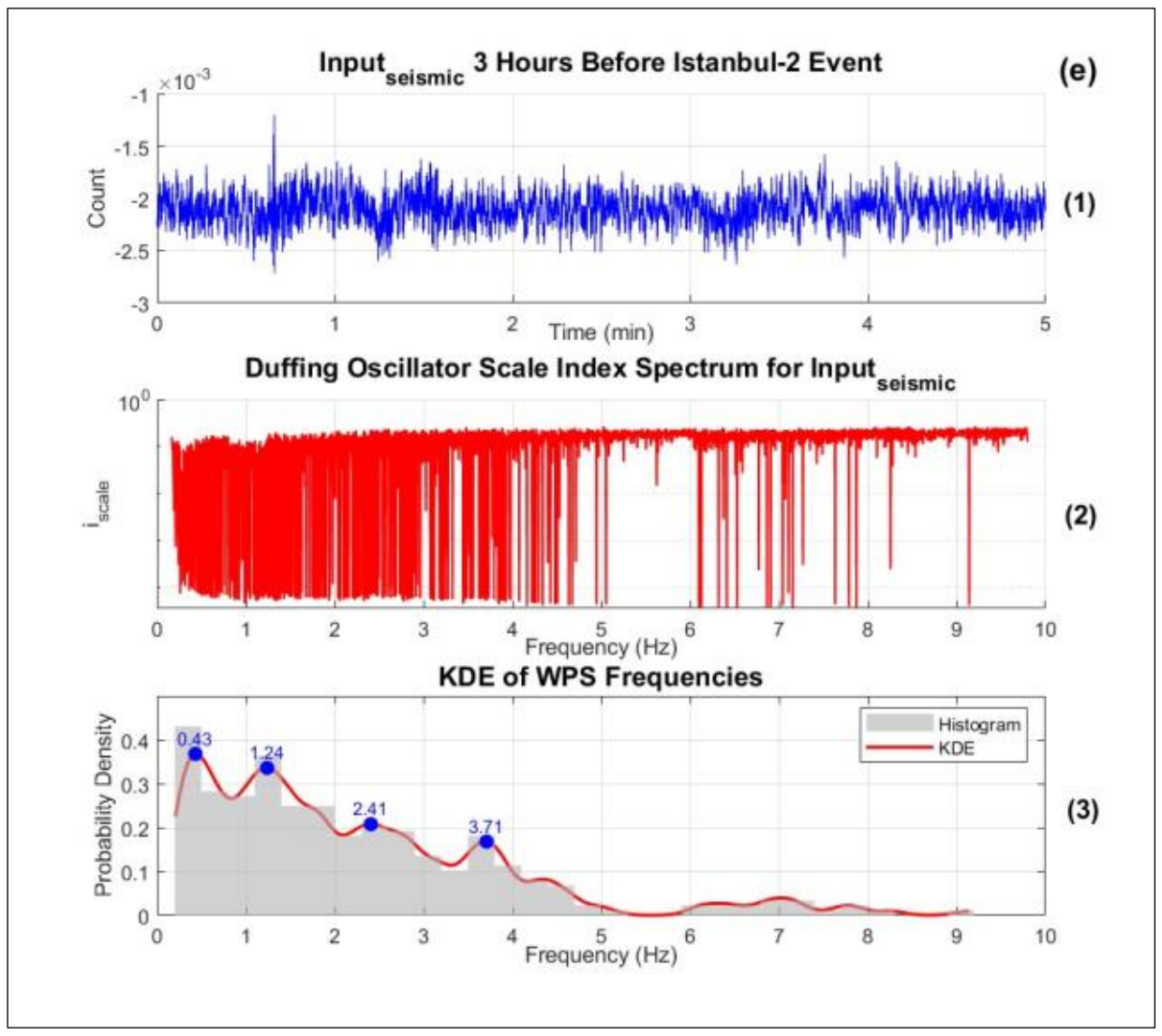

**Figure 4:** Comparison of seismic input, Duffing oscillator wavelet scale index spectrum, and WPS frequency distributions across five time windows relative to the Istanbul-2 event. Panels (a–e) correspond to: (a) during/after the event (5-min), (b) 0–5 min before, (c) 1 hour before, (d) 2 hours before, and (e) 3 hours before. For each window, subplots 1–3 show: (1) Seismic wave time-series, (2) Duffing oscillator wavelet scale index spectrum, and (3) KDE of WPS peak frequencies. Together, the 15 subplots (a1–e3) illustrate the evolution of seismic signals, nonlinear spectral behaviour, and dominant frequency content leading up to and during the event.

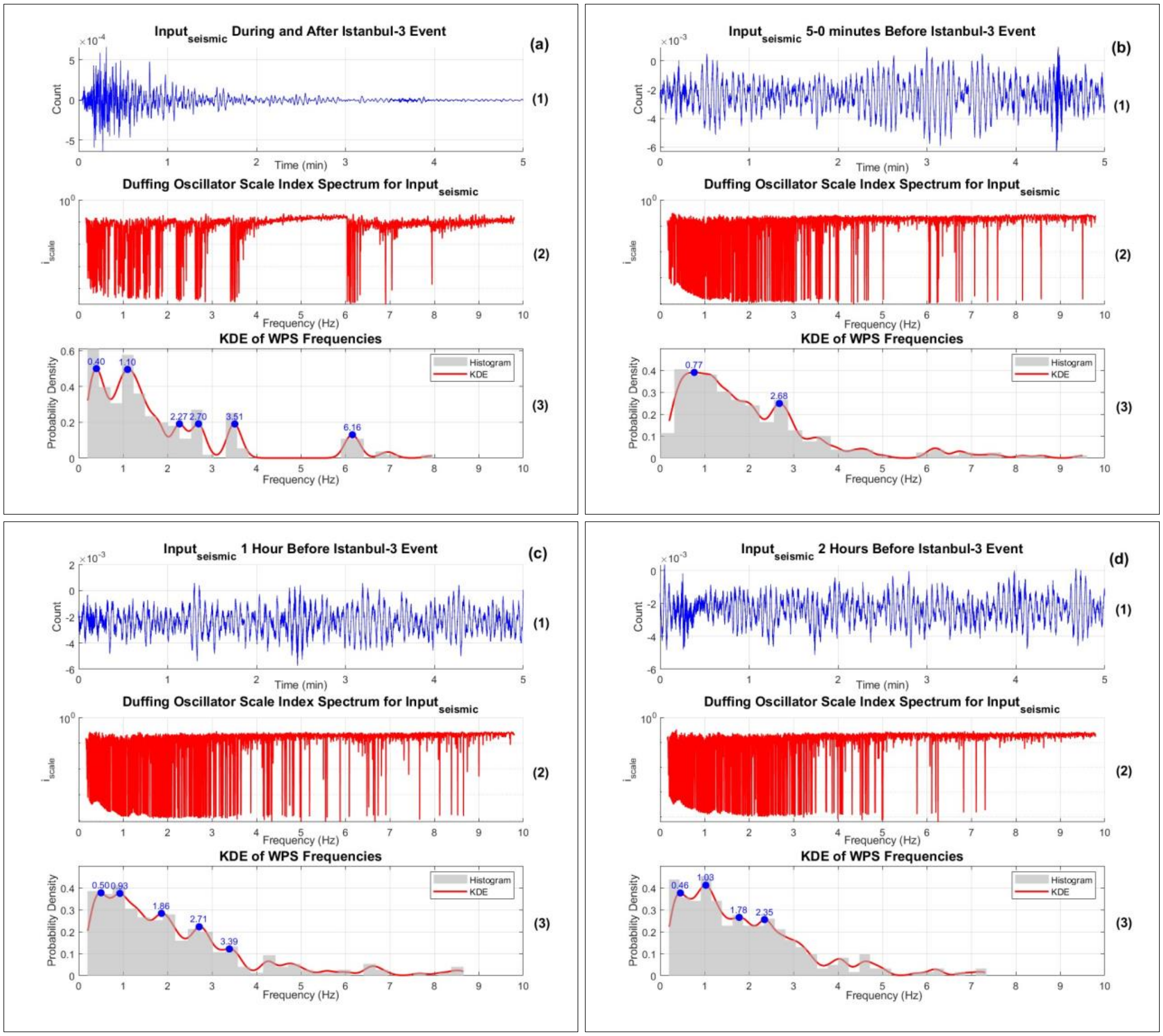

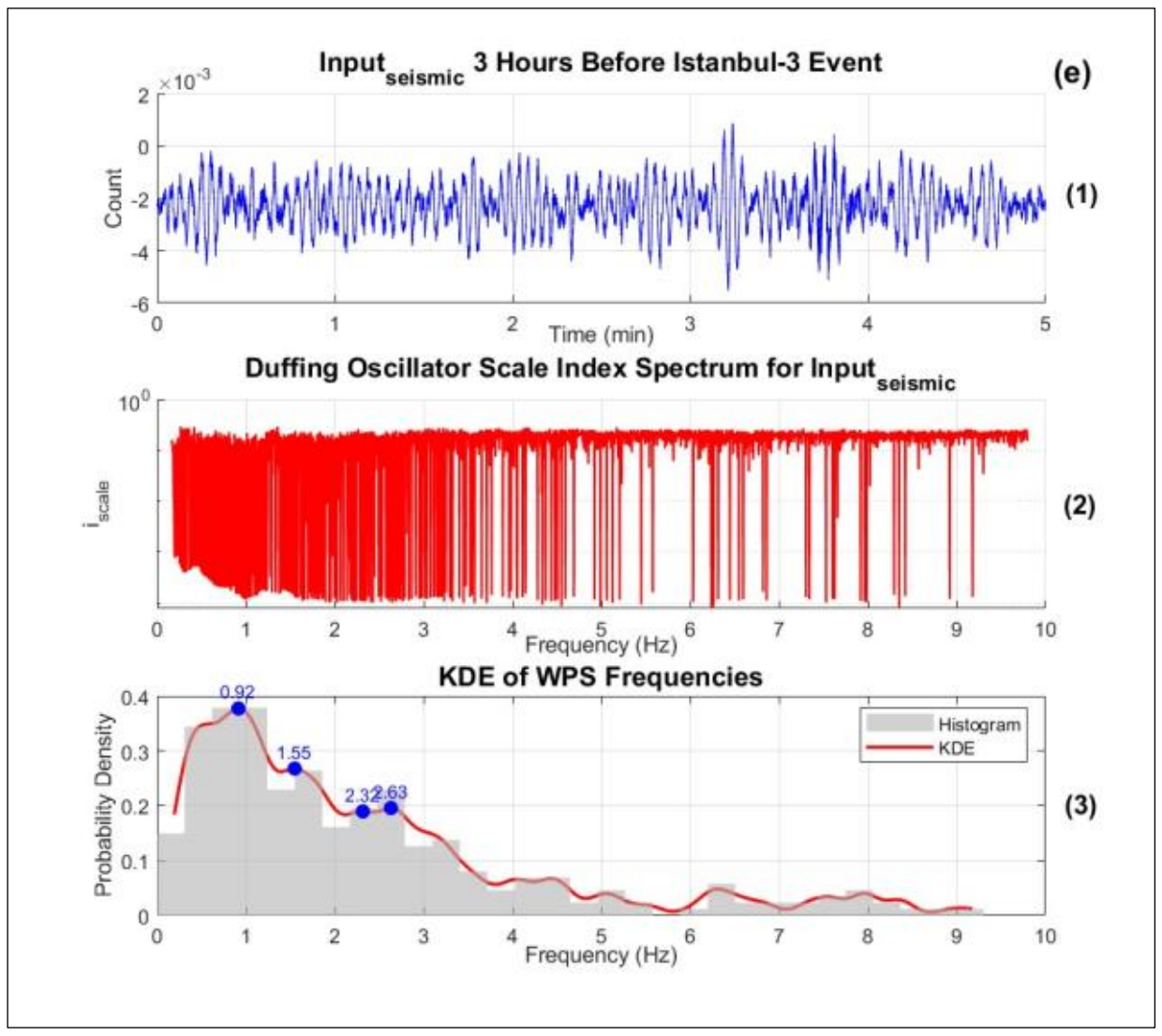

**Figure 5:** Comparison of seismic input, Duffing oscillator wavelet scale index spectrum, and WPS frequency distributions across five time windows relative to the Istanbul-3 event. Panels (a–e) correspond to: (a) during/after the event (5-min), (b) 0–5 min before, (c) 1 hour before, (d) 2 hours before, and (e) 3 hours before. For each window, subplots 1–3 show: (1) Seismic wave time-series, (2) Duffing oscillator wavelet scale index spectrum, and (3) KDE of WPS peak frequencies. Together, the 15 subplots (a1–e3) illustrate the evolution of seismic signals, nonlinear spectral behaviour, and dominant frequency content leading up to and during the event.

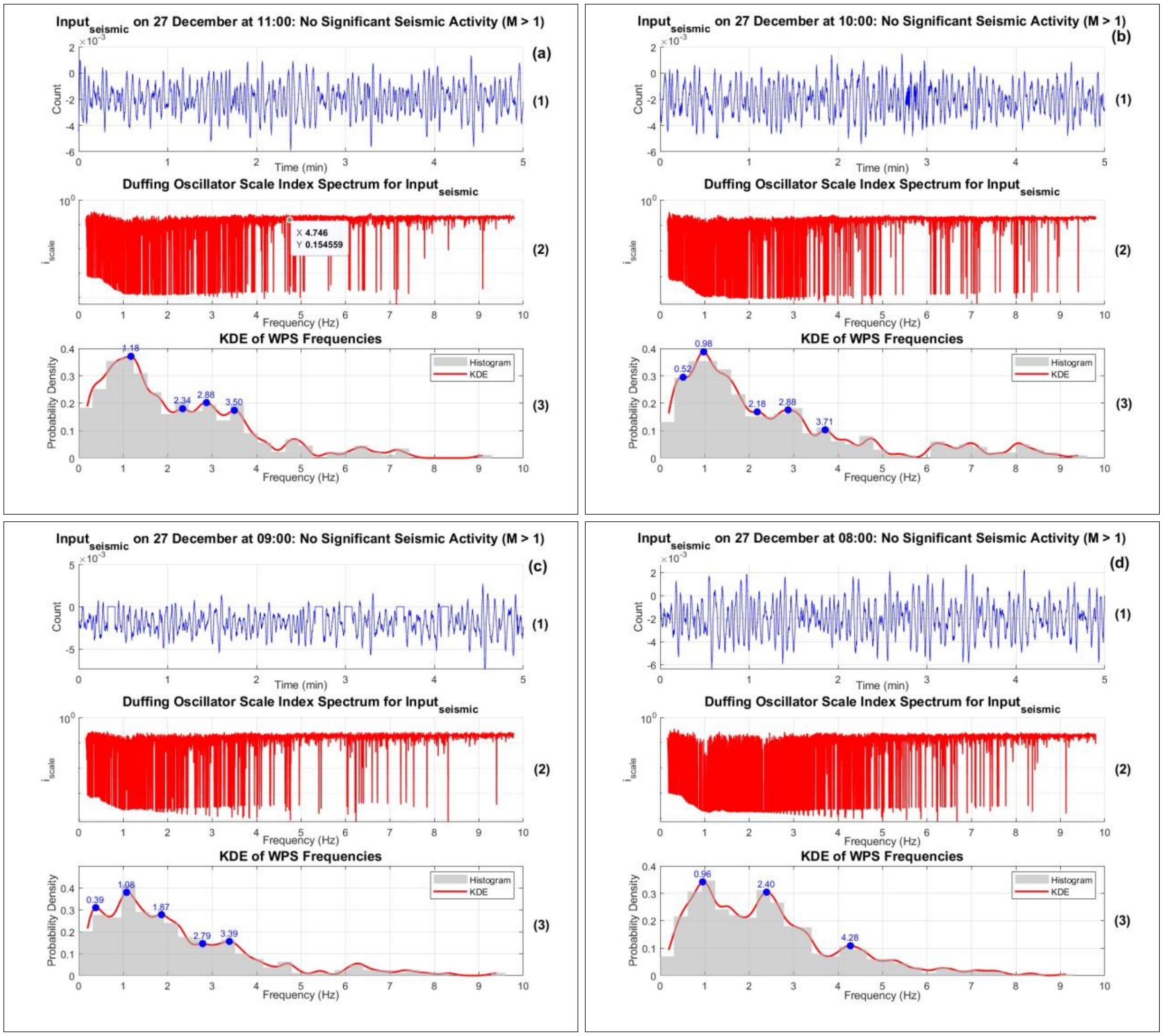

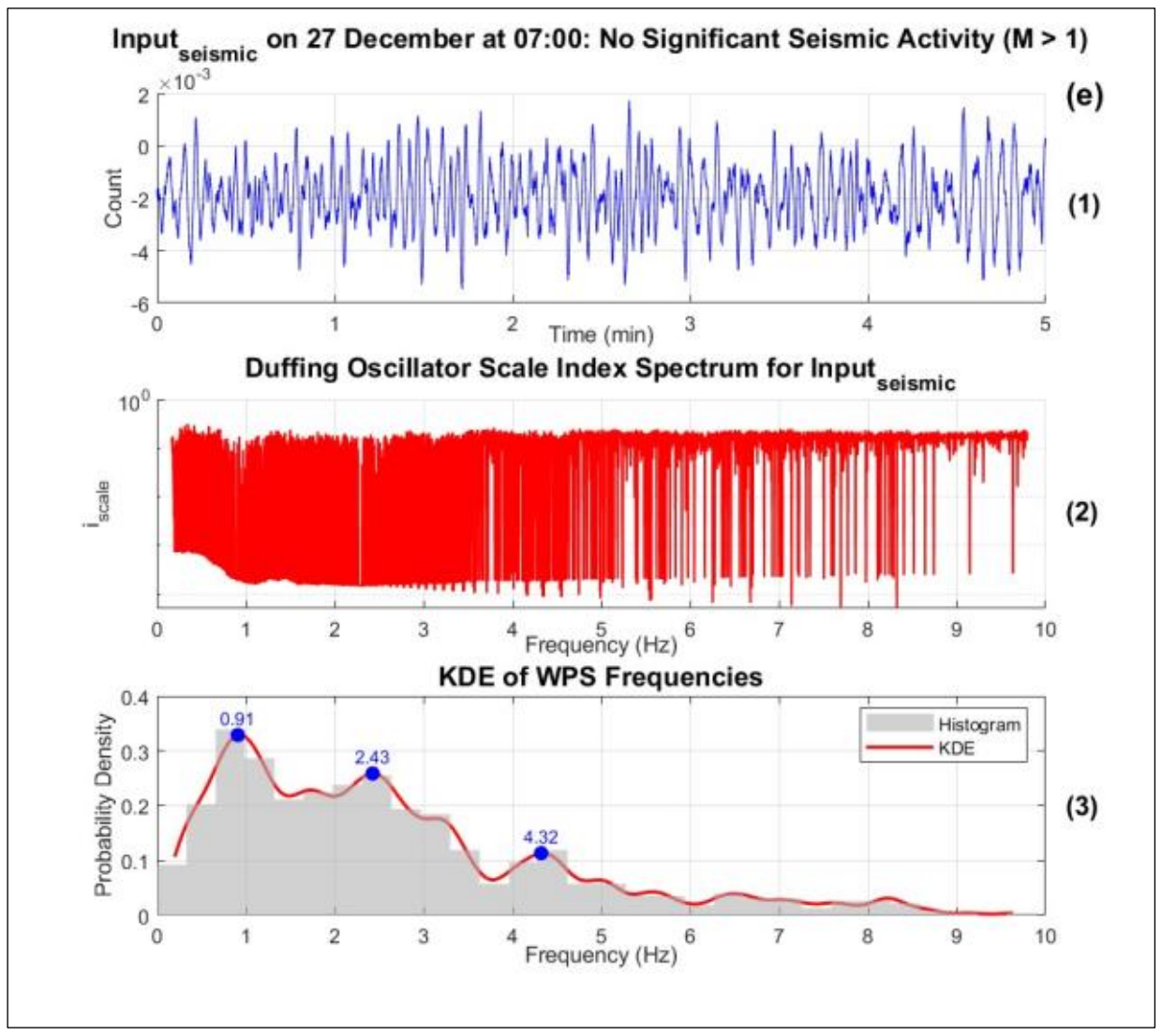

**Figure 6:** Comparison of seismic input, Duffing oscillator waveelt scale index spectrum, and WPS frequency distributions for control periods with **no significant seismic activity (M > 1)**. Panels **(a–e)** correspond to five separate 5-minute windows recorded under quiet seismic conditions. For each window, subplots **1–3** show: (1) $input_{\text{seismic}}$ time-series, (2) Duffing oscillator wavelet scale-index spectrum, and (3) KDE of WPS peak frequencies. Together, the 15 subplots (**a1–e3**) characterize baseline seismic behaviour and serve as a reference for comparison with earthquake-related analyses.

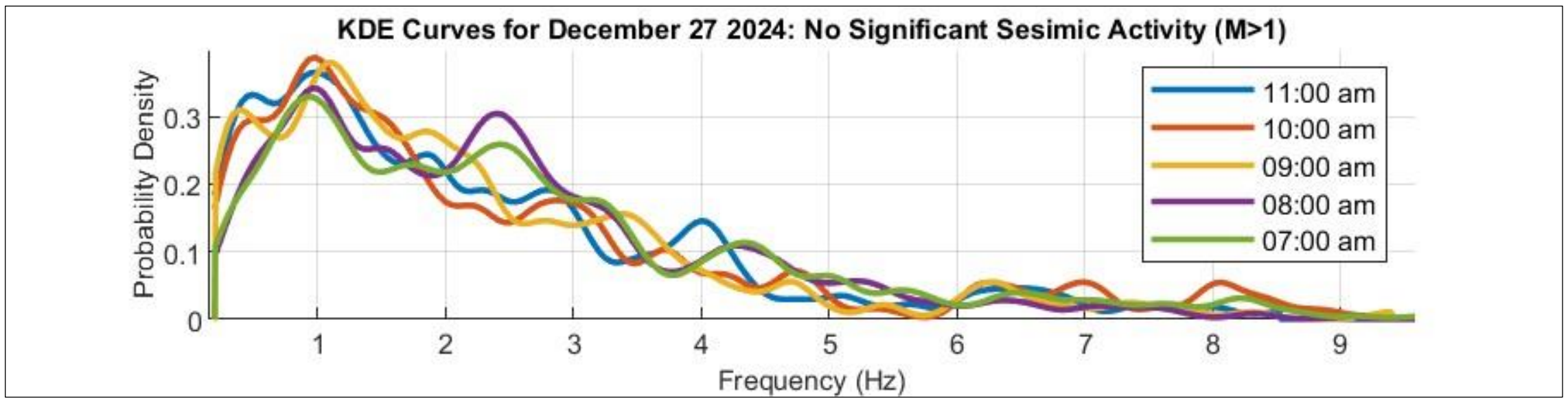

(a)

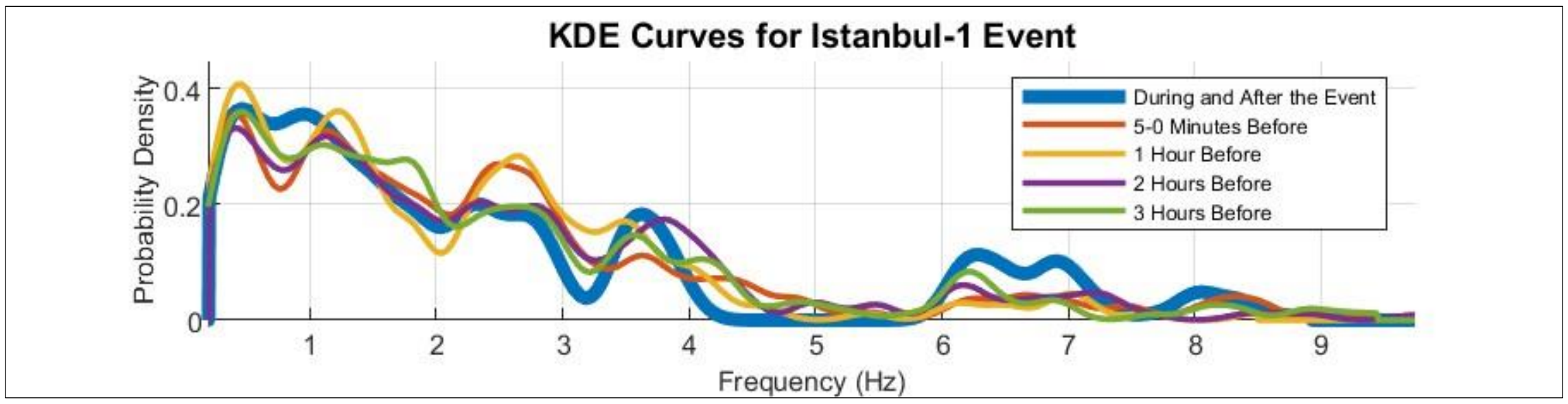

(b)

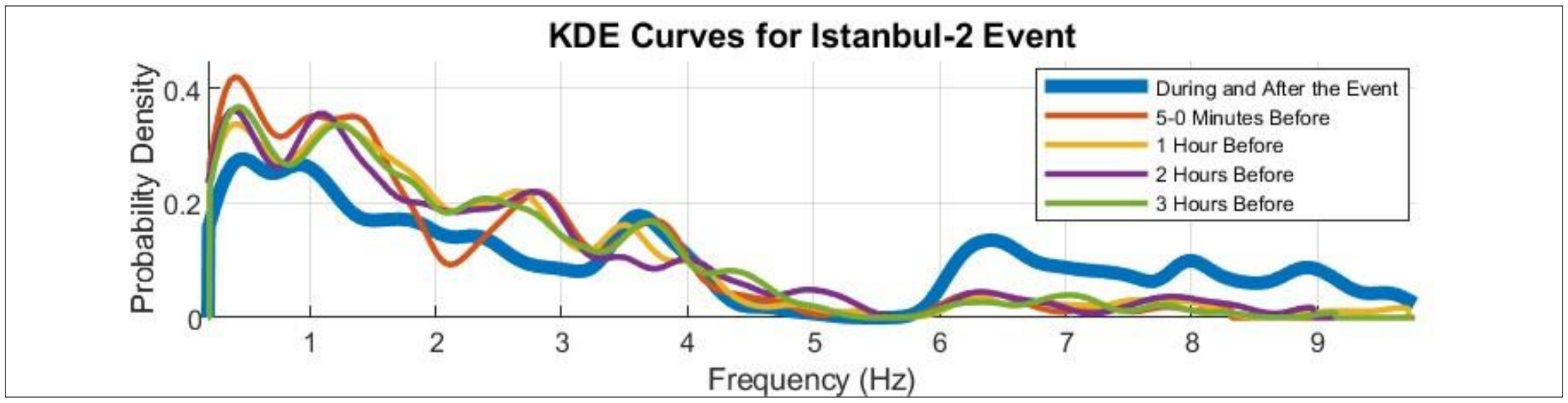

(c)

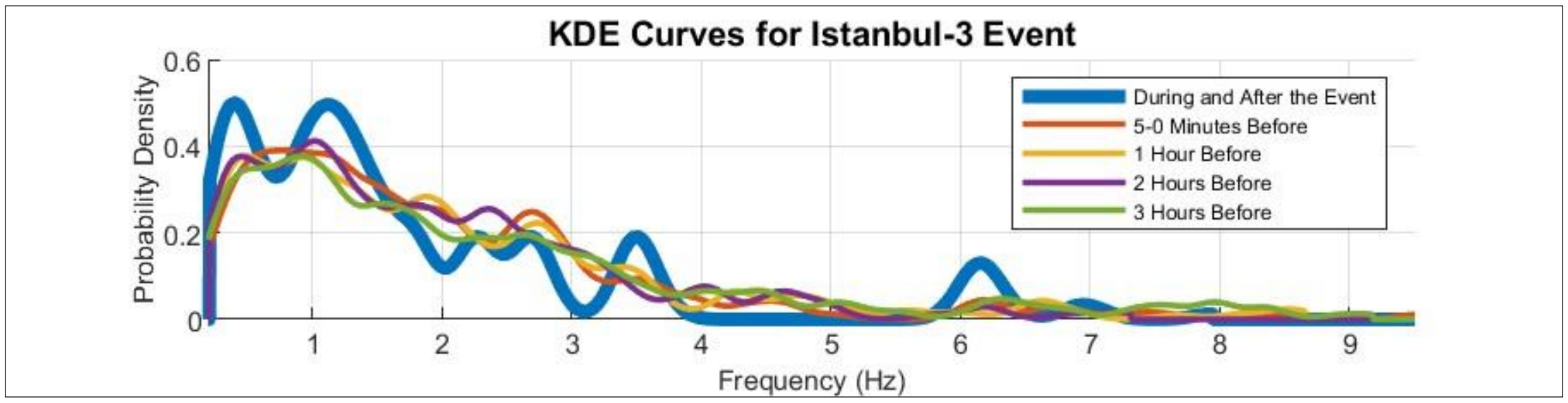

(d)

**Figure 7:** Kernel density estimate (KDE) curves of WPS peak frequencies for three earthquake events and one control period with no significant seismic activity. For each event graph, the five curves correspond to: during/after the event, 0–5 minutes before, 1 hour before, 2 hours before, and 3 hours before. In comparision, the control figure presents KDEs for five consecutive 5-minute windows spanning a quiet 5-hour period. The four graphs illustrate differences between seismic wave spectral behaviours.

Examination of the resulting frequency density revealed two principal outcomes. First, we consistently identified a stable and repeatable frequency-density peak near 0.5 Hz for all three events. This feature appeared not only during the co-seismic phase but also extended at four hours into the pre-seismic interval. On the other hand, the data representing seismic inactivity exhibited a peak around 1.0 Hz, indicating a different dynamical regime when no impending rupture was present. This shift from a 1.0 Hz quiescent peak to a 0.5 Hz pre-seismic data suggests a potential dynamical transition within the fault system as it approaches failure.

Second, the frequency-density distribution in the 5.0–9.8 Hz band remained low and relatively featureless during seismic inactivity and pre-seismic intervals but showed peaks during the earthquake events. This behaviour is consistent with the emergence of high-energy broadband motion characteristic of strong ground shaking, compared to the weak, deterministic periodic structure detectable before the fault rupture.

The time series consisting of low amplitude seismic waves recorded before the events were scaled by a factor of $2.5\times10^{-7}$ to bring their amplitudes into the operational range of the Duffing oscillator. Applying the inverse of this scaling yields the amplitudes of the detected weak periodic signals: $\gamma_{\mathrm{WPS}} \geq \frac{\gamma_c-\gamma}{2.5\times10^{-7}} \approx 0.4\text{–}200$ counts. For the seismic recordings during the events, a smaller gain coefficient of $2.5\times10^{-10}$ was required due to the substantially larger signal amplitudes. In this case, the identified weak-signal amplitudes satisfy $\gamma_{\mathrm{WPS}} \geq \frac{\gamma_c-\gamma}{2.5\times10^{-10}} \approx 400\text{–}200{,}000$ counts. To quantify the strength of the weak signal relative to the seismic background, the RMS values of the seismic datasets: $\mathrm{RMS} = \sqrt{\frac{1}{N}\sum_{i=1}^{N} x_i^2}$ N=30,000 samples per 5-minute window and the Seismic-to-Weak-signal Ratio ( $\mathrm{SWR_{dB}}$ ) was computed: $\mathrm{SWR_{dB}} = 10\log_{10}\frac{0.5\gamma_{\mathrm{WPS}}^2}{\mathrm{RMS}^2}$.

Pre-seismic data: RMS ≈ 2.55–2.79 $\times10^4$ counts, $SWR_{dB} \approx -79.9$ to $-79.3$ dB;
During the seismic event: RMS ≈ 1.55–2.10 $\times10^6$ counts, $SWR_{dB} \approx -57.4$ to $-54.8$ dB.

The results show that the Duffing oscillator scale index method is efficient at very low SNR value of ≈ -80 dB. However, when the weak periodic component is deeply buried within noise, with even lower SNR, the ability of the oscillator to detect the transition from chaotic to periodic state decreases, which is a limiting factor of the oscillator [25]. Furthermore, the adjustment of the gain coefficient and the damping parameter δ used in input scaling the the Duffing oscillator play important roles in the WPS detection. Deviations from optimal parameter values, as well as background seismic noise or other external effects, can decrease the oscillator sensitivity to weak signals. Our findings also highlight the significance of this careful parameter tuning and thorough calibration for effective application in earthquake monitoring systems.

## 5. Discussion

In this work, we applied a novel methodology for detecting and classifying weak periodic signals based on wavelet scale index analysis of the Duffing oscillator. We adapted the method explicitly for characterising the low amplitude seismic records before and during three seismic events and during an inactive period. We were able to detect deterministic components embedded within seismic recordings by exploiting the nonlinear sensitivity of the Duffing system near the chaos and periodic state boundary. For this, we used data from three moderate-to-strong earthquakes, as well as a inactive control period along the NAF.

This automated methodology showed clear advantages for WPS detection over conventional phase-space diagram analysis with its computational efficiency and suitability for real-time seismic monitoring. This approach provided a scalable and robust framework for identifying WPS that would otherwise remain hidden in seismic signals. It also showed strengths compared to stochastic resonance and chaotic resonance as it allowed WPS analysis without the need for noise optimization or the simplification of target waveforms into pure sinusoids. This work further improved the Scale index-Duffing based automated WPS detection [26], using a single scan to obtain density information across a wide frequency spectrum from low amplitude seismic wave time series.

We were able to detect frequency-density peak near 0.5 Hz during and at pre-seismic intervals before the events. However, the data obtained during seismic inactivity exhibiting a peak around 1.0 Hz, indicating a different dynamical behaviour. This shift from a 1.0 Hz quiescent peak to a 0.5 Hz pre-seismic data suggests a potential dynamical transition within the fault system as it approaches failure. We also demonstrated that the frequency-density distribution in the frequency range of 5.0–9.8 Hz remained low during pre-seismic intervals and during inactive period but showed peaks during the earthquake events. This behaviour can explain the emergence of higher frequency motion characteristic of strong ground shaking.

This work allows the emergence of a possible follow up researches. First, extending the analysis to longer pre-seismic windows beyond four hours will help determine the onset time at which the quiescent 1.0 Hz signature transitions into the emergent 0.5 Hz precursor. Establishing this transition point can be important for evaluating the reliability of these spectral features as earthquake precursors. Second, while the methodology is largely automated, further refining the numerical pre-adjustments such as the input gain coefficient and frequency dependent amlitude parameter will improve robustness across diverse stations, magnitudes, and sampling rates. These advancements will potentially enable large-scale implementation of this Duffing based WPS detection method in regional early-warning systems and deepen our understanding of the nonlinear processes governing complex seismic phenomena.

In conclusion, this nonlinear methodology of density estimation of weak periodic signals has been proved to be a robust tool for pre-seismic signal analysis. The study also emphasizes that this improved nonlinear weak periodic signal detection methodology can be implemented to extract hidden behaviours in other various fileds such us physiology or finance to study big data with chaotic characteristics.

**Declaration of competing interest**

The author declares that there are no competing interests relevant to the content of this article.

**Data availability Statement**

The datasets generated during and/or analysed in this article are available from the author on reasonable request.

**Funding Statement**

No funding was received to support this work.

**Acknowledgment**

I would like to thank Prof. Dr. K. Gedi Akdeniz and Dr. Mahmut Akıllı for their valuable discussions on the methodology and for their helpful comments on the manuscript.